\documentstyle[12pt]{article}
\newcommand{\sect}[1]{\setcounter{equation}{0}\section{#1}}
 \begin{document}
\def\bq{\begin{equation}}
\def\eq{\end{equation}}
\begin{flushright}
{UT-KOMABA-96/20}\\
{\sl  October, 1996 }\\
\end{flushright}

\begin{center}
{\large\bf DIAGRAMMATIC ANALYSIS OF THE TWO-STATE QUANTUM HALL SYSTEM 
WITH CHIRAL INVARIANCE
}
\end{center}

\begin{center}
{\bf  S. Hikami and K. Minakuchi$^{ }$} 
\vskip 2mm
{$^{ }$ Department of Pure and Applied Sciences, University of Tokyo}\\
{Meguro-ku, Komaba, Tokyo 153, Japan}\\
\end{center}

\vskip 5mm
\begin{abstract}
 The quantum Hall system in the lowest Landau level with Zeeman term 
is studied by 
a two-state model, 
which has a chiral invariance. Using a diagrammatic analysis,
we examine this two-state model with random impurity scattering, 
and find the exact value of the conductivity 
at the Zeeman energy $E = \Delta$. We further study the conductivity 
at the another extended state $E = E_1$ ($ E_1 > \Delta$).
We find that the values of the conductivities at $E = 0$ and $E = E_1$
do not depend upon the value of the Zeeman energy $\Delta$. 
We discuss  also the case where the Zeeman energy $\Delta$ 
becomes a random field.
\end{abstract}
\vskip 5mm
\newpage
\sect{Introduction}

 The critical behavior around the extended state  in the two dimensional 
quantum Hall system 
has been studied by various methods. Recently, the spin degenerate 
case attracted interests. The spin-up state and the spin-down state 
almost degenerate when the Zeeman energy is small. It is considered that 
 these two states can be mixed by the impurity scattering.
 
 One of author,  Shirai and Wegner{\cite{HSW}}(HSW) considered a 
two-state 
 model in the lowest Landau level, in which the impurity scattering occurs only 
between different spin states. This model corresponds to the strong 
spin-orbit scattering limit, in which the spin should be changed at each 
impurity scattering.
Remarkably there appear three extended states 
in this model,
one is at the band center $E = 0$, and the other two at $E = \pm E_1$.
The conductivity at $E = 0$ has been obtained exactly by a diagrammatic 
analysis and becomes $\sigma = e^2/2\pi^2 \hbar$\cite{HSW}. This 
model 
has a chiral invariance; the energy eigenvalues always appear in 
the positive and negative pairs. Then the  state at $E = 0$ 
becomes a special state, 
which can hybridizes itself due to this chiral invariance.{\cite{DKKLee}}
The density of state near $E = 0$ is not broden so much by the 
impurity scattering. 
Therefore, the density of state at $E = 0$ is enhanced and could be 
singular. 
The $E = 0$ state is a resonant state. 
At $E = 0$, all scattering effects are 
remarkably 
cancelled out for the conductivity, and the localization effect is smeared out.
 This cancellation occurs not only for 
the Gaussian white noise distribution but 
also for the  general local non-Gaussian random distribution.{\cite{HSW}}

 This  model has been examined further 
by the numerical method\cite{Hanna,Minakuchi} and the localization exponents
have been estimated for these extended states.
The localization length exponent at $E = 0$ seems different from the usual 
quantum Hall system and belongs a new universality class, although other 
two extended states at $E = E_1$  belongs to the conventional quantum Hall 
universality class with the localization length exponent $\nu 
\simeq  2.3$
\cite{Minakuchi}.

The state at $E = 0$ in this model has been suggested to 
be relevant to the 
 chiral Dirac fermion model with a  random 
vector potential  \cite{Ludwig}, which gives a singularity for the density of 
state. The value of the conductivity for this random vector potential model
agrees with the value of HSW model.

 In the previous paper\cite{Minakuchi}, the Zeeman term has been included,
 which does not break a chiral invariance.
It has been shown that 
 the density of state has a gap less than the Zeeman energy $\Delta$,
and the extended state shifts from $E = 0$ to the Zeeman energy $E = \Delta$. 

In this paper, we further consider this extended HSW model with a Zeeman term 
by a diagrammatic method. We evaluate  the exact value of the longitudinal 
conductivity at $E = \Delta$.
Also we will discuss the extended state at $E = \pm E_1$ ($E_1 > \Delta$),
 which is 
believed to belong the conventional quantum Hall universality class.
We argue that the value of the conductivity at $E = E_1$ becomes same as 
the conductivity for no-Zeeman case $\Delta = 0$ by the diagrammatic analysis.
This  is consistent with the numerical result{\cite{Minakuchi}}.
We show exactly that the inclusion of the Zeeman term does not 
alter the values of the conductivities of the 
extended state. This result may be expected but we verify it by a 
diagrammatic expansio method.
When the Zeeman energy becomes 
a random variable, the situation will be changed. 
We briefly discuss this random 
Zeeman energy case by the diagrammatic method.
\vskip 5mm
\sect{ Diagrammatic analysis of the two-state{\hskip 2mm}
 quantum Hall system}
\vskip 5mm

   The Hamiltonian for the two-spin state may be described by $2\times 2$
matrix\cite{Minakuchi}
\bq\label{2.1}
 H = {1\over{2m}}( p - e A)^2 + \left(\matrix{\Delta & v^{\dag}(r)\cr
        v(r)&-\Delta\cr}\right)
\eq 
where $v^{\dag}(r)$ and $v(r)$ are random potentials at the spacial point $r$.
 The constant $\Delta$ represents the Zeeman energy.
In the Landau quantization, the up-spin state and the down-spin state acquires 
the Zeeman energy $\pm \Delta$.
 The matrix of the second term of (\ref{2.1})
acts on the spin state, which eigenstate 
is represented by a vector of two components. 
The distribution of 
these random potential $v(r)$ is  assumed as a Gaussian white noise 
distribution, i. e. 

\bq\label{2.2}
      < v(r)>_{av} = < v^{\dag}(r) >_{av} = 0
\eq
\bq\label{2.3}
      <v^{\dag}(r) v(r^{\prime}) >_{av} = w \delta( r - r^{\prime})
\eq

The diagrammatic expansions for the one particle Green function and 
the two-particle Green function for the lowest Landau level has been 
investigated \cite{Hikami1,Hikami2,Hikami3}. In the case 
of no-Zeeman term, a useful expansion for the diffusion constant $D$ was 
derived, 
 and indeed by this expression, the exact value of the conductivity 
was obtained\cite{HSW}. Note that we mainly consider the Gaussian white 
noise ditribution in this paper but the exact evaluation of the conductivity is also 
applied to any local non-Gaussian random potential as shown in \cite{HSW}.

In the two dimensional case, the Green function for the lowest Landau 
level is simply expressed by

\begin{eqnarray}\label{2.4}
   G(r) &=&<< r | {1\over{E - H}}| r>>_{av}\nonumber\\
        &=& {1\over{A_1 + i A_2}}
\end{eqnarray}
where G(r) has a traslational invariant, and $A_1$ and $A_2$ become 
real numbers independent of $r$. This is due to the quantization under 
a strong magnetic field. For example, the density of state $\rho(E)$ 
becomes simply as $- A_2/ \pi (A_1^2 + A_2^2)$.

When the two-spin state model is considered, we have two different Green functions 
$G_A(r)$ and $G_B(r)$. The notation A and B are the spin-up state and 
the spin down state, respectively. 
Using the self-energy $\Sigma$ for A and B, we obtain by definition,

\bq\label{2.5}
  A_1 = 2\pi (E - {1\over{2}}\hbar \omega_c - \Delta - {1\over{2\pi}}
{\rm Re} \Sigma_A )
\eq
\bq\label{2.6}
  A_2 = 2\pi ({\epsilon\over{2}}  - {1\over{2\pi}}{\rm Im} \Sigma_A ) 
\eq

\bq\label{2.7}
  B_1 = 2\pi ( E - {1\over{2}} \hbar \omega_c + \Delta - {1\over{2}}
{\rm Re} \Sigma_B )
\eq
\bq\label{2.8}
  B_2 = 2\pi ({\epsilon\over{2}} - {1\over{2\pi}}{\rm Im} \Sigma_B )
\eq

The diagrammatic expansion follows the previous studies and the convenient 
method for obtaining the coefficients of each orders may be found in 
\cite{Hikami1,Hikami2}.
As the first exercise, let us approximate the self-energy $\Sigma$ by 
the Green function itself.
Then we have $\Sigma_A = 2\pi w /(B_1 + i B_2)$ and $\Sigma_B = 2\pi w /
(A_1 + i A_2)$.

It may be convenient to represent two Green functions by 
 $G_A = C_A e^{i\theta_A}$, 
$G_B = C_B e^{i\theta_B}$, and also $x = C_A C_B$.
From (\ref{2.6}) and (\ref{2.8}), in the limit $\epsilon \rightarrow 0$,
we obtain $\theta_A = \theta_B$, $x = 1$. We represent the energy  
$E - {1\over{2}}\hbar \omega_c$ simply by $E$.

From (\ref{2.5}) and (\ref{2.7}), using $x^2 =4\pi^2 w^2/(A_1^2 + A_2^2)(B_1^2 
+ B_2^2) = 1$, we obtain
\begin{eqnarray}\label{2.9}
   A_1 &=& 2\pi ( E - \Delta ) - {1\over{2\pi w}} 
         B_1 ( A_1^2 + A_2^2)\nonumber\\
       &=& 2\pi ( E - \Delta ) - {1\over{w}} 
           ( E + \Delta ) ( A_1^2 + A_2^2 ) + A_1
\end{eqnarray}
Thus we obtain $A_1^2 + A_2^2 = 2\pi w (E - \Delta )/( E + \Delta)$.
Similary, we get $B_1^2 + B_2^2 = 2\pi w ( E + \Delta)/( E - \Delta )$.

Then (\ref{2.5}) becomes
\bq\label{2.10}
   A_1 = 2\pi ( E - \Delta ) - B_1 {(E - \Delta)\over{(E + \Delta)}}
\eq

From (\ref{2.6}), we have $A_2 = 2\pi w B_2/(B_1^2 + B_2^2) = B_2 (E - \Delta)
/(E + \Delta)$. Further noting that $A_1/A_2 = B_1/B_2$, and from (\ref{2.10})
we obtain the following solution,
\bq\label{2.11}
     A_1 = \pi ( E - \Delta )
\eq
Similary we get 
\bq\label{2.12}
     B_1 = \pi ( E + \Delta )
\eq

The imginary parts $A_2$ and $B_2$ are obtained from $A_1^2 + A_2^2 =
2\pi w (E - \Delta)/(E + \Delta)$. They become
\bq\label{2.13}
   A_2 = {1\over{2}}\sqrt{ {E - \Delta\over{E + \Delta}}}
   \sqrt{4 w - (E^2 - \Delta^2)}
\eq
\bq\label{2.14}
   B_2 =  {1\over{2}}\sqrt {{E + \Delta\over{E - \Delta}}}
   \sqrt{4 w - (E^2 - \Delta^2)}
\eq
The density of state $\rho_A$ and $\rho_B$ are given by the $\rho_A = - 
A_2/\pi (A_1^2 + A_2^2)$, $\rho_B = - B_2/\pi (B_1^2 + B_2^2)$.
Since $A_1^2 + A_2^2 = 2\pi w (E - \Delta)/( E + \Delta)$, $B_1^2 + B_2^2 = 
2\pi w ( E + \Delta)/( E - \Delta )$, the density of state $\rho(E)$ has a gap 
between $-\Delta < E < \Delta$, and has the inverse square root singurality 
at $E = \pm \Delta$.
This behavior resembles to the density of state of the superconductor.
Note that we fix the Zeeman energy parameter $\Delta$. Later, we consider 
the average over this $\Delta$ for the density of state.

We now go beyond this approximation by expanding  the self-energy in the power series of $w$. In this two-state 
model with the zeeman term, the diagrams become same as the two-state model 
without Zeeman term \cite{HSW}. Using the notations $A_2/A_1 = - \tan \theta_A$
and $B_2/B_1 = - \tan \theta_B$, and $x = C_A C_B (2\pi w)$, $G_A = C_A e^{i\theta_A}$
and $G_B = C_B e^{i\theta_B}$ ($-\pi/2 < \theta_A < 0$, $ -\pi/2 < \theta_B
< 0$), we have
\begin{eqnarray}\label{2.15}
  {\pi \epsilon\over{A_2}} &=& 1 - x {\sin \theta_B\over{\sin \theta_A}}
- {1\over{4}}x^3 {\sin ( 3 \theta_B + 2 \theta _A) \over{ \sin \theta_A}}
- {2\over{5}} x^4 {\sin( 4 \theta_B + 3 \theta_A )\over{ \sin \theta_A}}
\nonumber\\
&+& \cdots.
\end{eqnarray}
 
\begin{eqnarray}\label{2.16}
      {\pi\epsilon\over{B_2}} &=& 1 - x {\sin \theta_A\over{\sin \theta_B}}
- {1\over{4}}x^3 {\sin ( 3 \theta_A + 2 \theta_B) \over{ \sin \theta_B}}
- {2\over{5}} x^4 {\sin( 4 \theta_A + 3 \theta_B )\over{ \sin \theta_B}} 
\nonumber\\
&+&\cdots.
\end{eqnarray}
Up to order $x^{10}$, the expansion coefficients are given in (4.1) of 
Ref.{\cite{HSW}}.

At $E = \pm \Delta$, the phases $\theta_A$ and $\theta_B$ become $- { \pi\over{2}}$.
This is evident within the first order 
approximation by (\ref{2.12}) and (\ref{2.14});
 $A_2/A_1 = \tan \theta_A = 0$ and 
$B_2/B_1 = \tan \theta_B = 0$. Beyond this order, it remains also true. We have evaluated the real part of the Green function numerically by the same method 
of Ref.\cite{Minakuchi}, and find that the real part vanishes at $E = \pm 
\Delta$. 

The conductivity in the lowest Landau level is obtained from Kubo-formula
by the diagrammatic expansion\cite{Hikami3}. As an equivalent method, we 
have Einstein relation $\sigma = e^2 D \rho$, where $D$ is a diffusion 
constant. Here we use Einstein relation, since a diagrammatic expansion 
is simpler\cite{Hikami1}.
The diffusion constant $D$ is defined as the coefficient of $q^2$ in
the inverse of the two-particle correlation function $K(q)$;

\bq\label{2.17}
     K(q) = \int <<r| {1\over{E - H + i0}}|r^{\prime}><r^{\prime} | {1\over{
E - H - i 0}} | r >>_{av} e^{-iq( r - r^{\prime})}d^2r^{\prime}
\eq

This $K(q)$ is expanded in the power series of w. The Feynman rule for this 
expansion may be seen in the previous literatures\cite{Hikami1}. We have 
for the small momentum $q$,
\bq\label{2.18}
     {K(q = 0)\over{K(q)}} = 1 + {D\over{\epsilon}} q^2
\eq
Since we have two different propagator $G_A$ and $G_B$, the two-particle 
correlation function $K(q)$ is also devided into two parts, $K_A(q)$ 
and $K_B(q)$. And the diffusion constant $D$ also is defined differently by 
(\ref{2.18}). The denominator $\epsilon$ in (\ref{2.18}) can be expressed 
by (\ref{2.15}) and (\ref{2.16}).
Then finally, we obtain the following equations which are the modification 
of the previous expression\cite{HSW}.

\bq\label{2.19}
    {2\pi D_B\over{B_2}} = 1 - {x^3\over{4}}( \cos ( 2\theta_B + 2\theta_A ) +
    \cos (\theta_A + \theta_B)) + \cdots
\eq
\bq\label{2.20}
    {2\pi D_A\over{A_2}} = 1 - {x^3\over{4}}( \cos ( 2\theta_B + 2\theta_A ) +
    \cos (\theta_A + \theta_B)) + \cdots
\eq
The imaginary part of $G_A$ and $G_B$ are propotional to the density of state $\rho$. The conductivity $\sigma_{xx}$ is given by Einstein relation,
\bq
    \sigma_{xx} = {1\over{2}}( e^2 {D_A} \rho_A + e^2 {D_B} \rho_B )
\eq
At $E = \Delta$, and $E=-\Delta$, we have 
$A_1 /A_2 = B_1 /B_2 = 0$ as explained before.
Thus, we have $\theta_A = \theta_B = -\pi/2$. Remarkably all corrections 
cancells out in (\ref{2.19}) and (\ref{2.20}) except one. 
These cancellations are essentially 
same as the previous case without Zeeman term\cite{HSW}.
The conductivity $\sigma_{xx}$ at $E = \pm \Delta$ 
becomes $e^2/2\pi^2 \hbar$, which is same value for the no-Zeeman term at $E=0$.

Thus, we have found the exact value of the longitudinal conductivity 
at $E = \pm \Delta$. In the previous numerical work\cite{Minakuchi}, this value was 
obscure, although it suggested the similar value.
We find no particular difference for the conductivity 
between the extended state at $E = \Delta$ 
and  the $E = 0$ in the $\Delta = 0$ case.

 The effect of our Zeeman term on the density of state can be also discussed 
by the matrix model. As an  analogous matrix model to HSW model, a complex 
block matrix model has been studied and the universal oscillation of the 
density of state near $E = 0$ has been obtained in the large N limit, 
where $N$ is a size of the matrix\cite{HZ,BHZ}. The simple matrix model is 
given by
\bq
     M = \left ( \matrix{ \Delta & v^{\dag}\cr
                          v & - \Delta }\right )
\eq
where $\Delta$ is a unit matrix multiplied by $\Delta$, and $v^{\dag}$ is 
 a $N\times N$ complex matrix.  
It is a straighforward exercise to evaluate the density of state for the 
finite N through Kazakov method\cite{BHZ,Kazakov}, since this matrix model has a chiral 
invariance; the eigen values appear always in a pair of positive and negative 
one. The effect of this Zeeman term $\Delta$ is just a shift of the 
energy $E$. When we take the large N limit first in this model, the 
density of state coincides with (\ref{2.13}) and (\ref{2.14}). However, 
there is a crossover to the oscillatory behavior near $\Delta$ in the 
small region of order $1/N$\cite{BHZ}.
\vskip 5mm

\sect{ Extended state at $E = E_1$ }

As pointed out by the numerical works\cite{Hanna,Minakuchi},
there are extended states at $E  = \pm E_1$, which is greater than
$\Delta$.
 It was suggested that the conventional 
universality class of the quantum Hall effect with a localization exponent $\nu 
\simeq  2.3$ is realized at $E = E_1$ \cite{Hanna,Minakuchi}. The shift 
of the energy from $E = 0$ to $E = E_1$ is due to the effective magnetic 
field effect due to the off-diagonal random potential $v$.

The shift of the conventional extended state of quantum Hall system to 
$E = E_1$ has been observed by several models.  The  Chalker-Coddington 
network model\cite{Chalker} was extended to include 
 the spin-scattering, 
 and the shift of the extended state is shown 
with the same localization exponent \cite{LeeChalker,Wang,Horovitz}.

Since the previous work \cite{HSW}
 did not discuss this
extended state at $E = E_1$, we first consider this state for the no-Zeeman
term $\Delta = 0$ case.    
The diagrammatic expansion for $D/A_2$ was given up to order$ x^8$\cite{HSW}.
The series for the diffusion constant $D$ without Zeeman term becomes

\begin{eqnarray}\label{3.1}
{2\pi D\over{A_2}} &=& 1 - {1\over{4}} ( \cos 4\theta + \cos 2\theta )x^3 
- (0.32 \cos 6\theta + 0.16 \cos 4\theta + 0.16 ) x^4 \nonumber\\
&-& (1.14279155188 \cos 8\theta + 0.715564738292 \cos 6\theta \nonumber\\
&+& 0.180555555555 \cos 4\theta + 0.75195133149 \cos 2\theta\nonumber\\
&+& 0.144168962351 ) x^5
\nonumber\\
&-&( 4.01604212958 \cos 10\theta + 2.10780216729 \cos 8\theta \nonumber\\
&+& 0.228564968429 \cos 6\theta + 1.65837390674 \cos 4\theta\nonumber\\
&+& 0.613624866859 \cos 2\theta + 1.09205589084 ) x^6
\nonumber\\
&-& (16.8938594252 \cos 12\theta + 8.85669612784 \cos 10\theta\nonumber\\
&+& 1.34798158141 \cos 8\theta + 5.49180725809 \cos 6\theta \nonumber\\
&+& 1.75117610591 \cos 4\theta + 6.74855019206 \cos 2\theta \nonumber\\
&+& 1.10403646547 ) x^7\nonumber\\
&-& ( 79.7915118420 \cos 14\theta + 40.5552408026 \cos 12\theta \nonumber\\
&+& 5.99939335079 \cos 10\theta + 20.196866445487\cos 8\theta \nonumber\\
&+& 4.4215378158747 \cos 6\theta + 23.475132715585 \cos 4\theta \nonumber\\
&+& 6.4926520588331 \cos 2\theta + 12.477855103819 ) x^8 
+ \cdots.
\end{eqnarray}
where the variable $x$ is solved by the asymptotic expansion of (\ref{2.15}).
Putting $\theta_A = \theta_B$, $\epsilon = 0$, we have up to the third 
order of $x$, 
\bq\label{3.2}
   x \simeq 1 - {1\over{4}} {\sin 5 \theta \over{\sin \theta}}
\eq
This approximation shows the maximum of $x$ at $\theta \simeq - 0.9$. 
The maximum value of $x$ becomes 1.3. The value of $x$ becomes zero 
for $\theta \rightarrow 0$ from (\ref{2.15}).
 
This is quite similar to the case of the conventional quantum Hall case: 
the exact value of x at the band center is $x = 4/\pi = 1.2732$ \cite{Weg}.
And $x$ becomes zero for $\theta \rightarrow 0$.
Thus the point $\theta = -0.9$ for this two-state quantum Hall system 
corresponds to the band center of the one-state quantum Hall system.
The shift appears due to the off-diagonal two-state random potential.
Up to order $x^3$, from (\ref{3.1}), we obtain by inserting the value of x,
\bq\label{3.3}
   {2\pi D\over{ A_2}} = 1 - {1\over{4}} ( \cos 4\theta + \cos 2 \theta ) 
( 1 - {1\over{4}} {\sin 5 \theta \over{\sin \theta}} )^3
\eq
The maximum of $2 \pi D/A_2$ becomes 1.6 at $\theta = -0.9$.
The conductivity $\sigma$ is obtained by multiplying $e^2 \sin^2 \theta
/2 \pi^2 \hbar$ to the value of  
$2\pi D/A_2$.
We have analysed up to order $x^3$. We think the maximum peak of the 
conductivity remains finite for the higher order analysis.
And also we think that the state at $\theta = -0.9$ corresponds to the 
band center of the one-state quantum Hall system, and becomes 
extended. This is consistent 
with the previous numerical result \cite{Minakuchi}, which shows there is 
an extended state at $E = E_1$ except $E = 0$. The states of the energy 
$ 0 < E < E_1$ and $E > E_1$ are considered to be localized. For the 
investigation of the localization, we need the renormalization group 
analysis via $1/N$ expansion \cite{Hikami1}, and we do not discuss it 
here. 

For the Zeeman case ($\Delta \ne 0$), the series of (\ref{3.1})
 is modified as (\ref{2.19}) 
and (\ref{2.20}), where $2 \theta$ is replaced by $\theta_A + \theta_B$.
In general, $\theta_A$ is not equal to $\theta_B$. The range of these angles 
are between $-\pi/2$ and $0$. We assume that there is an extended state  
at $E = E_1$ for 
the Zeeman case. Then, we find that if $\theta_A + \theta_B$ is 
same as the critical value $\theta_c$ in (\ref{3.1}), we have the same expression
for (\ref{2.19}) and (\ref{2.20}). Since there is one extended state, we 
have $\theta_A = \theta_B$ at $E = E_1$.
This is a duality between A-state and B-state. 
Then we find that the same conductivity as the no-Zeeman case at $E = E_1$.
The conductivity is obtained from (\ref{2.19}) by multiplying a factor 
$\sin^2 \theta$, which is $A_2^2/(A_1^2 + A_2^2)$ for the case $A_1 \ne 0$.
Indeed our previous numerical result shows this behavior.
This argument of the equivalence does not determin the absolute value of 
the conductivity, but it verifies that the value of the conductivity 
at $E = E_1$ does not depend upon the Zeeman energy $\Delta$.
\vskip 5mm
\sect{Random Zeeman energy model}
\vskip 5mm

In the previous sections, we assumed that the Zeeman energy $\Delta$ is 
a fixed constant. 
 When this $\Delta$ in (\ref{2.1}) is a random field, which depends upon 
the spacial coordinate $r$, 
the situation becomes different. The distribution of this random field 
$\Delta(r)$ is Gaussian.
We will discuss this random Zeeman energy model 
 by a diagrammatic expansion method.

Instead of the Zeeman energy $\Delta$, we represent it by 
a random field $u(r)$.
Then the second term of (\ref{2.1}) becomes
\bq\label{4.1}
V(r) = \left(\matrix{ u(r) & v^{\dag}(r)\cr
            v(r) & - u(r)\cr }\right )
\eq
where $r$ is a place of the impurity scattering.
This model represents the spin flip at $r$ due to the random field $v$ and 
the random Zeeman energy by $u(r)$. There is no correlation between 
$v(r)$ and $u(r)$. The matrix $V(r_1)$ does not commute with the matrix $V(r_2)$. We have to 
consider the successive operation of the random scattering at $r_1, r_2, 
\cdots, r_N$ for the eigenstate. The eigenstate is represented by a vector of two components.
The random variable $u$ has the following average,
\bq
   < u(r) u(r^{\prime}) >_{av} = w^{\prime} \delta (r - r^{\prime})
\eq
Then the diagrammatic expansions of (\ref{2.15}) and (\ref{2.16}) become
a series of the scattering strength $w$ and $w^{\prime}$. Note that some 
terms have a negative sign due to the minus sign in (\ref{4.1}) in the 
matrix element.

In this random Zeeman energy model, the chiral invariance is broken.
The scattering appears  between a state A 
and a state B but also between the same spin state due to 
the diagonal random field $u(r)$. 

We find that after 
the average of the multiplication of the matrix $V(r_i)$ over 
the random distribution, the non-vanishing diagrams can be expressed  
by assigning the indecis A and B for the Green function. 

The self-energy of $\Sigma_A$ becomes by the diagrammatic expansion,
\begin{eqnarray}\label{4.3}
\Sigma_A &=& w^{\prime}  G_A + w G_B  - w w^{\prime} G_A G_B^2 \nonumber\\
&-& {1\over{4}}
( {w^{\prime}}^3 G_A^5 + w^3 G_A^2 G_B^3 + 3 w {w^{\prime}}^2
 G_A^2 G_B^3 )\nonumber
\\
&-& {1\over{3}} ( 3 w^3 G_A^5 + 3 w {w^{\prime}}^2 G_A^2 G_B^3 
- 2 w {w^{\prime}} 
G_A G_B^4 + w^{\prime} w^2 G_A G_B^4\nonumber\\
&-& 4 {w^{\prime}}^2 w G_A^3 G_B^2 + 2 w^{\prime} w^2 G_A^3 G_B^2 )
+ \cdots
\end{eqnarray}

From this equation, we have 
\begin{eqnarray}\label{4.4}
{\pi \epsilon\over{A_2}} &=& 1 - {1\over{\sin \theta_A}} ( w^{\prime} 
C_A^2 \sin \theta_A + w C_B C_A \sin \theta_B )\nonumber\\
&+& w w^{\prime} C_A^2 C_B^2 {\sin (\theta_A + 2 \theta_B)
\over{\sin \theta_A}}\nonumber\\
&-& {1\over{4}} ( {w^{\prime}}^3 C_A^6 {\sin  5\theta_A \over{\sin \theta_A}} 
+ (w^3 + 3 {w^{\prime}}^2 w ) C_A^3 C_B^3 
{\sin ( 2 \theta_A + 3 \theta_B )\over{ \sin \theta_A }})\nonumber\\
&-& {1\over{3}} [  3 w^3 C_A^6 {\sin 5 \theta_A \over{\sin \theta_A}} + 
3 w {w^{\prime}}^2 C_A^3 C_B^3 {\sin (2\theta_A + 3 \theta_B)\over{\sin \theta_A}}
\nonumber\\
&+& ( w^{\prime} w^2 - 2 w {w^{\prime}}^2 ) 
C_A^2 C_B^4 {\sin (\theta_A + 4 \theta_B )\over{ \sin \theta_A}}
\nonumber\\
&+& ( 2 w^{\prime} w^2 - 4 {w^{\prime}}^2 w ) C_A^4 C_B^2 
{\sin ( 4 \theta_A + 2 \theta_B )\over{ \sin \theta_A}} ]\nonumber\\
&+&  \cdots
\end{eqnarray}

By the symmetry between $A$ and $B$ states, we are able to put $\theta_A$ = 
$\theta_B$, $C_A = C_B$. Then, (\ref{4.4}) becomes simpler. 

It may be interesting to consider the three different cases: 1) $w^{\prime} << 
w$, 2) $w^{\prime} \sim w$, 3) $w^{\prime} >> w$. The case 1) corresponds to 
the  2-state model, which we have discussed previously as $\Delta = 0$.
The perturbation of the parameter $w^{\prime}$ can be 
obtained. 
The case 2) shows the strong effect of the random field $u(r)$. 
When, for example, $w^{\prime} = {1\over{2}} w$, the series of (\ref{2.15}) 
has alternative sign, and when $\theta_A = \theta_B = -\pi/2$ at $E = 0$, 
the density of state is suppressed.
This behavior is  similar 
to the gap state for the non-vanishing Zeeman 
energy $\Delta$, for which we have discussed in the section 2.
The case 3) is similar to the conventional quantum Hall sysytem, since 
two state $A$ and $B$ can be decoupled completely in the limit 
$w \rightarrow 0$. 
The extended energy $E = E_1$ approaches to $ E = 0$.

The case 1) can be studied by the perturbation of $1/N$. We need to generalize 
the model to the N-orbital model. The random field $u(r)$ in (\ref{4.1}) is
changed to $u(r)\times {\rm I}$ where ${\rm I}$ is $N\times N$ unit matrix.
$v$ is also a complex $N \times N$ matrix. 
The density of state in the $1/N$ expansion shows the logarithmic 
singularity at order $1/N^2$  for $w^{\prime} = 0$\cite{HSW}.
. In the lowest order of 
$w^{\prime}$, (\ref{2.15}) becomes at $E = 0$,
\begin{eqnarray}
   {\pi \epsilon\over{A_2}} = 1 -  w C^2 -  {w^{\prime}\over{N}} ( {w C^2\over{
1 - w C^2}} )  + 
{d_1\over{N^2}} \ln^2 ( 1 - w C^2 )
\nonumber\\
&+& \cdots
\end{eqnarray}
The $1/N$ term is obtained from the diagramms of Fig. 4. 
Up to order of $1/N$, solving the equation, we obtain
\bq
     w C^2 = 1 + {w^{\prime}\over{ 2 N w}} \pm \sqrt{{w^{\prime}\over{N w}}}
\eq
Thus the logarithmic divergence is smeared out for small $w^{\prime}$, since 
$\ln^2 \epsilon$ is changed to $\ln^2 w^{\prime}$.
 
In the presense of $w^{\prime}$, the conductivity $\sigma = e^2/2 \pi^2 \hbar$
is also changed. The logarithmic terms in the diffusion constant in $1/N^2$ 
order is cancelled by the vertex correction for $w^{\prime}$, 
indeed each diagram cancells 
completely not only for the logarithmic term, as we have 
seen in (\ref{3.1}).
When $w^{\prime} \ne 0$, this cancellation by the vertex part is not complete 
at $E = 0$, and the logarithmic term exists. 
Then, there appears  a localization for $E = 0$ when $w^{\prime} \ne 0$.
By the same reason, $E\ne 0$ state is also localized.
\vskip 5mm

\sect{ Discussion }

   In this paper, we have evaluated the exact value of the conductivity 
of the two state model with the fixed Zeeman term 
at the Zeeman energy and find that this result is consistent with the 
previous numerical result. We also observed that the conductivity 
at $E = E_1$ for the case of non-vanishing Zeeman energy is  
same as the conductivity at $E = E_1$ for the no-Zeeman energy. 
Thus the effect of the Zeeman term does not alter the values of the 
conductivities at $E = \Delta$, $E = E_1$.

 We discussed further how the situation is modified when the Zeeman energy 
becomes a random field, which obeys the Gaussian white noise distribution. 
Then, the diagrammatic expansion becomes two parameters $w$ and $w^{\prime}$. 
We find, in the first order of $w^{\prime}$, the cut-off of 
the singularity of 
 the density of state 
appears, and it leads to the localization at $E = 0$.

\vskip 5mm

{\bf Acknowledgement}
\vskip 3mm

This work is supported by a Grant-in-Aid for Scientific Reseach by the 
Ministry of Education, Science and Culture, and by CREST of Japan 
Science and Technology Corporation.


\end{document}